\documentclass[useAMS,usenatbib,usegraphicx]{mn2e}

\title[Low Mass X-Ray Binaries in M31 GCs] 
  {WFCAM Survey of M31 Globular Clusters: Low Mass X-ray Binaries}
\author[M. Peacock et al.]
{Mark B. Peacock$^{1}$\thanks{E-mail:m.b.peacock@phys.soton.ac.uk (MBP)},
Thomas J. Maccarone$^{1}$, Christopher Z. Waters$^{2}$, 
\newauthor
Arunav Kundu$^{2}$, Stephen E. Zepf$^{2}$, 
Christian Knigge$^{1}$, David R. Zurek$^{3}$\\
$^{1}$School of Physics and Astronomy, University of Southampton, Southampton, SO17 1BJ, UK\\
$^{2}$Department of Physics and Astronomy, Michigan State University, East Lansing, MI 48824, USA\\ 
$^{3}$Department of Astrophysics, American Museum of Natural History, New York, NY 10024, USA} 

\begin{document}

\date{Released 2008 Xxxxx XX}

\pagerange{\pageref{firstpage}--\pageref{lastpage}} \pubyear{2008}

\maketitle

\label{firstpage}

\begin{abstract}

We investigate the relationship between Low Mass X-ray Binaries (LMXBs) and globular clusters (GCs) using UKIRT observations of M31 and existing \textit{Chandra}, \textit{XMM-Newton}, and \textit{ROSAT} catalogues. By fitting King models to these data we have estimated the structural parameters and stellar collision rates of 239 of its GCs. We show a highly significant trend between the presence of a LMXB and the stellar collision rate of a cluster. The stellar collision rate is found to be a stronger predictor of which clusters will host LMXBs than the host cluster mass. We argue that our results show that the stellar collision rate of the clusters is the fundamental parameter related to the production LMXBs. This is consistent with the formation of LMXBs through dynamical interactions with little direct dependence on the neutron star retention fraction or cluster mass. 

\end{abstract}


\section{Introduction}

Since early studies of X-ray sources in the Milky Way it has been known that GCs are a rich source of LMXBs \citep[e.g.][]{Clark}. To explain the populations observed in the Milky Way \citep[][and references within]{Liu}, its GCs need to be two orders magnitude more efficient at forming LMXBs than other Galactic regions. Theorists have long argued that the likely reason for this is that GC LMXBs are formed mainly through dynamical interactions instead of the evolution of primordial binaries \citep[e.g.][]{Clark,Fabian,Sutantyo,Hills,Verbunt}. The dynamical processes which can form LMXB systems include the neutron star capturing a donor star through tidal capture; exchange interactions; and direct collisions. These interactions are likely to be very common in the cores of GCs where the stellar densities are relatively high. Studies of the 13 bright LMXBs in the Milky Way GCs are consistent with dynamical formation scenarios. 

In the \textit{Chandra} era it has been possible to study much greater numbers of LMXBs by looking at extra-galactic X-ray sources. Observations of nearby galaxies generally show large numbers of LMXBs associated with their GCs \citep[e.g. the review of][]{Fabbiano}. From these studies it has been possible to identify relationships between the properties of a cluster and the probability that it will contain a LMXB. 

In both the Milky Way and nearby galaxies it has been found that LMXBs favour brighter (and hence more massive) GCs \citep[e.g.][]{Sarazin,Kim06,Smits,Kundu07}. The likely reason for this is that higher mass clusters will generally have more stellar interactions and therefore form more LMXBs through dynamical interactions. It has also been suggested that higher mass clusters will retain more of the neutron stars they produce. This is because more massive clusters will generally have higher escape velocities and neutron stars may be formed with large kick velocities. However it is possible for neutron stars to form with lower velocities via electron-capture supernovae \citep[e.g.][]{Pfahl,Ivanova}. 

If dynamical formation is the primary method of forming LMXBs in GCs, we expect there to be a strong correlation between the stellar collision rate of a cluster and the presence of a LMXB. The stellar collision rates in GCs will be dominated by collisions in their cores (where the stellar densities are highest). Unfortunately, due to their small sizes and large distances from us, resolving the cores of extra-galactic GCs requires very good spatial resolution. Some previous work has tried to infer collision rates without resolving the cluster cores \citep[e.g. in M31;][]{Bellazzini}, but direct relationships between LMXBs and collision rate have been limited to GCs in the Milky Way and a few nearby galaxies observed with the \textit{Hubble Space Telescope (HST)}. \textit{HST} observations have been used to estimate the structural parameters of GCs around M31 \citep{Barmby02,Barmby07}, Cen~A \citep{Harris06,Jordan,McLaughlin} and M87 \citep{Jordan04}. From these studies correlations between stellar collision rates and the presence of LMXBs have been proposed in Cen~A and M87. 

In this letter we study 15 LMXBs in the M31 GC system. These are the closest extra-galactic GCs and the only ones for which reliable core radii can be estimated using ground based photometry. 

\section{Data}

To investigate the properties of the GCs in M31, we obtained K band photometry using the Wide Field CAMera (WFCAM) on the United Kingdom Infrared Telescope (UKIRT) under the service program USERV1652. These data cover most of the M31 disk, but avoid the central regions where surface brightness fluctuations are largest. The data include 239 of the confirmed M31 GCs. 

Each observation consisted of microstepped 5s exposures with a nine point jitter pattern. The 2$\times$2 microstepping was used to give an effective pixel size of 0.2$\arcsec$ and ensure the images were well sampled. Five observations were taken of each field to give a total exposure time of 225s. This gives a detection limit of K=19 at 5$\sigma$. 

The images were processed using the standard WFCAM pipeline \citep[for details on the WFCAM and its pipeline see e.g.][]{Dye}. The pipeline reduces and stacks the raw images and adds an accurate astrometric solution to the final images by fitting sources to the Two Micron All Sky Survey (2MASS). For combining our microstepped images we chose not to use the standard pipelines interleaving method. Instead the reduced images were drizzled together \citep{Fruchter} using the IRAF STSDAS task DRIZZLE. Drizzling has the advantage that it produces combined images with smoother PSFs than interlacing. 

To measure the structural parameters of the GCs observed, spatial resolution is very important. Our observations were taken on the nights of 2005-11-30 and 2007-08-06 with seeing of 0.85-1.00$\arcsec$ and 0.6-0.8$\arcsec$ respectively. At the distance of M31, this corresponds to a spatial resolution of 3.2-3.9pc and 2.1-2.9pc. This is better than can be achieved by \textit{HST} observations of Virgo Cluster galaxies where the spatial resolution is $\sim$8pc and only slightly worse than Cen~A where the resolution is $\sim$1.8pc. Our observations also have the advantage over these \textit{HST} observations that we are able to measure our PSF directly using stars in our fields instead of having to rely on models. 

\section{M31 Globular clusters}

\subsection{Selection of GCs}

Throughout this letter we select GCs using the Revised Bologna Catalogue (RBC) of M31 GCs v3.5 and adopt the naming conventions of this catalogue \citep{Galleti,Galleti06,Galleti07}. The RBC is based on the original catalogue of \citet{Battistini} and is regularly revised and supplemented to include new results on the GC system. We identify clusters in our images based on their locations in this catalogue. To minimise contamination of our sample we only include GCs present in the RBC which are marked as confirmed and which were identified from the work of \citet{Battistini,Battistini93,Auriere,Mochejska,Barmby00,Barmby02,Huxor}. In total 239 clusters are located in our observations with 115 present in multiple images. 

\subsection{Profile fitting} 

To investigate the structure of the GCs in our observations, we must first consider the effects of seeing on their appearance. The images of the GCs studied are the result of a convolution of their physical size with the PSF of the observations. To account for this, we create a model for the PSF of each image so that we can deconvolve it from the clusters and investigate their underlying properties. The PSF was modelled with a Moffat profile using the DAOPHOT tasks ALLSTAR/SEEPSF to fit the brightest 40 stars which were: unsaturated; had no bright neighbours; and were greater than 100 pixels from the detector edges (where the noise is significantly higher). No significant variation in the PSF was observed across the images so we select a single PSF model for each detector of each observation. 

To find the structural parameters of the GCs we fit spherical, single mass King models \citep{King} to their profiles using the program SUPERKING by Waters et al. (in prep.). The full details of this fitting and analysis will be presented in a subsequent paper. Here we present the results of the fits along with some consistency checks to test the reliability of the parameters found. The models were first convolved with the appropriate PSF for the image before being fit to the cluster profile. The best fitting model was then selected based on $\chi^{2}$ minimisation and the structural parameters for this model were output. 

Before fitting the GC profiles, bright stars (above 5$\sigma$) were removed from the GC region using the DAOPHOT task ALLSTAR. Each cluster was fit out to a radius of 20$\arcsec$. This was chosen to extend beyond the expected tidal radius for 85$\%$ of the clusters \citep[based on Milky Way GCs in the Harris catalogue;][]{Harris96}. This allows for the tidal radius and background level to be computed as accurately as possible. 

\subsection{Derived parameters}

\begin{table*}
 \centering
 \begin{minipage}{175mm}
  \caption{Catalogue of M31 GC parameters \label{GC data}}
\end{minipage}
 \begin{minipage}{150mm}
  \begin{tabular}{@{}lccccccccccc@{}}
  \hline
  \hline
   GC Name$^{(1)}$ & $\chi^{2}/\nu$ & W0 & c & K & r$_{c}^{(2)}$ & r$_{h}^{(2)}$ & r$_{t}^{(2)}$ & $log(\rho_{c})$ & $log(\Gamma)$ & Flag$\_$x$^{(3)}$ & Flag$^{(4)}$\\
   & & & & (mag) & (pc) & (pc) & (pc) & (L$_{\odot}$pc$^{-3}$) & & &\\
  \hline
   B001 & 1.07 & 6.25 & 1.32 & 13.7 & 1.25 & 2.71 & 25.9 & 3.87 & 6.00 & 0 & 1\\
   B003 & 1.15 & 9.55 & 2.25 & 14.9 & 0.24 & 4.05 & 42.7 & 4.82 & 5.99 & 0 & 1\\
   B005 & 1.20 & 8.25 & 1.91 & 12.5 & 0.36 & 2.27 & 29.1 & 5.57 & 7.47 & S01,T04 & 1\\ 
  \hline
  \end{tabular}\\
\end{minipage}
 \begin{minipage}{175mm}
 Table is available in full as Supplementary Material in the electronic edition of the journal or from the VizieR archive\\
 $^{1}$Names taken from the RBC of M31 GCs \citep{Galleti07} \\
 $^{2}$Assuming the distance of M31 to be 780kpc \citep{McConnachie} \\
 $^{3}$X-ray source associated with GC from; \citet{Trudolyubov} (T04); \citet{Supper} (S01); or no known source (0)\\
 $^{4}$Reliability flag based on visual examination of the cluster and its best fitting model. 
\end{minipage}
\end{table*}

Table \ref{GC data} shows the best fitting structural parameters for the 239 GCs studied. Where clusters were present and well fit in more than one observation, we select the observation with the best seeing. We assign a flag to each cluster based on visual examination of the cluster and its residual after subtraction of our best fitting model. These flags are used to identify poorly fitting models, as well as clusters contaminated by very bright stars. 

Using these parameters we estimate the stellar collision rate ($\Gamma$) for each cluster via $\Gamma \propto \rho_{c}^{3/2}r_{c}^{2}$. This relationship assumes a constant collision rate per unit volume in the cluster cores which dominates the total collision rate. It follows from the more general formula $\Gamma\propto\rho_{c}^{2}r_{c}^{3}/\sigma$ by assuming Virial equilibrium to estimate the velocity dispersion as $\sigma\propto\rho_{c}^{1/2}r_{c}$ \citep{Verbunt}. This relationship is used to calculate the collision rates presented in table \ref{GC data}. 

Despite studying only the confirmed clusters in the RBC it is possible that our sample contains some contamination from stellar sources. We identify misclassified clusters by finding objects whose stars should have collided many times over a Hubble time. We can estimate the timescale for stellar collisions from $\Gamma$ by assuming the cluster is comprised of solar type stars in order to find the constant in the above proportionality \citep[see eq. 3][]{Verbunt}. This gives that the timescale on which we expect a star to collide is:

\begin{equation}
 t_{coll} = \frac{N_{\star}}{\Gamma_{T}} = 1.1\times10^{9} \left(\frac{\rho_{c}}{\rm{pc}^{-3}} \right) \left(\frac{r_{c}}{\rm{pc}}\right)^{3} \left(\frac{\Gamma}{10^{5}\rm{yr}^{-1}}\right)^{-1}
 \label{eq-tcoll}
\end{equation}

$\Gamma_{T}$ is the total collision rate and $N_{\star}=(4/3)\pi r_{c}^{3}\rho_{c}$ is the total number of stars in the cluster core. From this equation we find 5 GCs to have $t_{coll}<10^{9}\rm{yr}$. One of the clusters (B431) has a bright star contaminating its profile. However the other four clusters (B006D, B096D, B292D, and B480) are well fit and we propose that they are likely to be misclassified stars. It has previously been suggested that B292D may not be a GC \citep{Huxor} and we note that the radial velocity estimates in the RBC of three of these clusters are consistent with being Milky Way halo stars. We remove these clusters from the following analysis. 

\subsection{Consistency checks} 

\begin{figure}
 \includegraphics[height=84mm,angle=270]{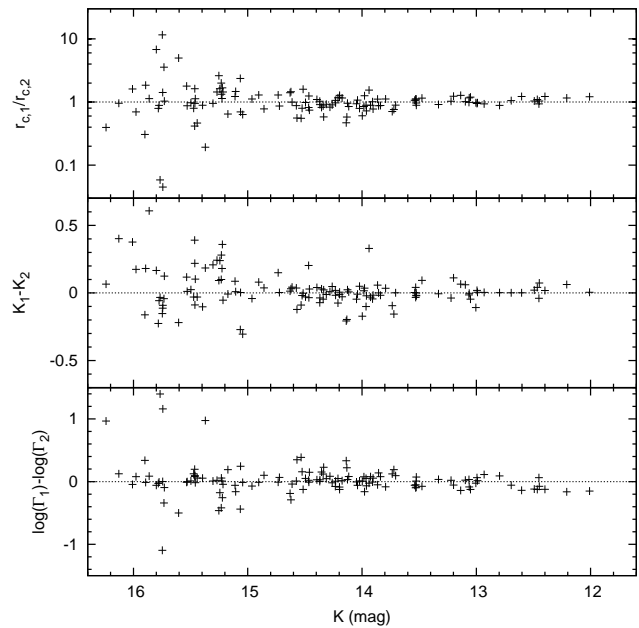}
 \caption{Comparison of the parameters derived from fitting the same cluster in different observations.}
 \label{fig:compare}
\end{figure}

Having identified the best fitting King model to describe each cluster we investigate the reliability of the parameters found. Due to the small spatial scale of the clusters we are fitting, it is likely that the errors on these parameters will be dominated by errors in the PSF model which is convolved with the cluster. 

To investigate the magnitude of the typical measurement errors, we compare the results obtained from fitting the same cluster from different observations. In total, 115 clusters are present in multiple images. These observations have all been taken under slightly different conditions and will have different PSF models. Therefore by fitting these clusters independently and comparing the resulting parameters we can estimate the reliability of the parameters calculated. Figure \ref{fig:compare} shows the differences between the derived parameters for the clusters fit in more than one image. 

For clusters brighter than K=15mag we find good agreement between the parameters found. However it can be seen that the scatter increases significantly for clusters with K$>$15mag. This suggests that the errors on the parameters found for these faint clusters are significantly higher. The large deviation in the magnitudes found for these faint clusters is likely to be due to contamination from nearby stars or surface brightness fluctuations. 

We expect that the variation observed in figure \ref{fig:compare} gives a reasonable estimate of the errors on the parameters found for all the GCs studied. In the following analysis, we include all 235 GCs studied but note that the errors on individual parameters for the faintest clusters will be large. Since most of the GCs studied and all X-ray GCs are brighter than this, our conclusions should not be sensitive to this increased error. 

\subsection{Comparison with previous work} 

\begin{figure}
 \includegraphics[height=84mm,angle=270]{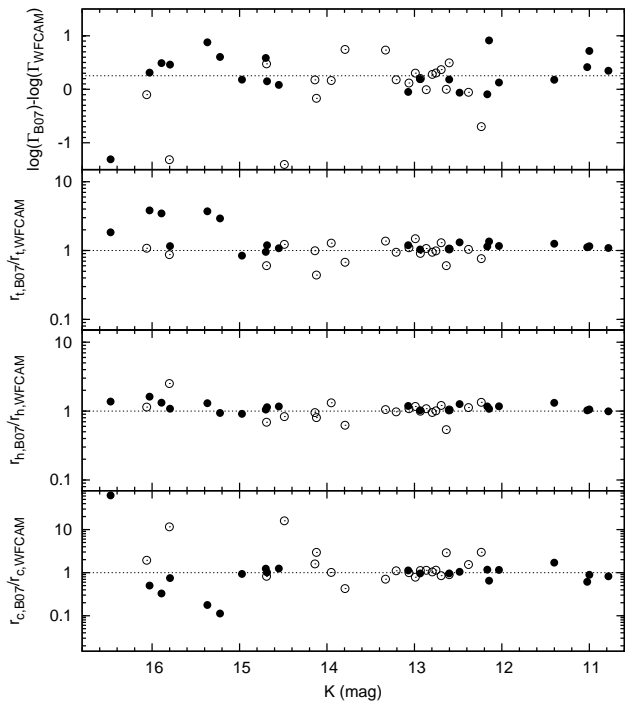}
 \caption{Comparison of the parameters measured here with those published by \citet{Barmby07} (B07) using \textit{HST} STIS/ACS (solid circles) and WFPC2 (open circles) observations.}
 \label{fig:compare-hst-ukirt}
\end{figure}

Some of the GCs in M31 have been observed by the \textit{HST} under several different programs. Using these data the structural parameters for 96 clusters have already been estimated from either ACS, STIS or WFPC2 observations \citep{Barmby02,Barmby07}. Figure \ref{fig:compare-hst-ukirt} compares these previously found parameters with those found in this study. There are 33 clusters present in both datasets. 

For clusters brighter than K=15mag, we find reasonable agreement between the radii found in this study and those found by \citet{Barmby07}. For clusters fainter than this we find significant differences between both the core and tidal radii found. However this is within the large errors predicted in the previous section for these fainter clusters. 

For all the cluster parameters compared we find greater discrepancies between our results and those based on WFPC2 observations (open circles). The likely reason for this is the methods used to estimate these parameters. For our data, we fit the whole profile of the cluster and select the best fitting tidal radius. While the WFPC2 observations are fit to the inner region of the cluster (where the signal to noise is greater) and the tidal radius inferred from the best fitting model \citep{Barmby02}. This second method is much more susceptible to slight deviations from a pure King model. We therefore believe our tidal radii may better represent the actual tidal radii of the clusters. 

The stellar collision rates found by both studies show a clear scaling difference. This difference is to be expected and is primarily because we estimate the number of stars based on K-band instead of V-band photometry. This offset is therefore most likely due to the colour of the clusters. Since our analysis is relative only to values measured in this study, this offset should not effect our conclusions. Also since stellar mass is more closely correlated with K-band luminosity than any optical band, comparisons of the collision rates within our data are likely to be robust. 

\section{X-ray sources in M31 GCs}

X-ray sources have been associated with GCs in M31 from \textit{ROSAT} observations \citep{Supper} and archived \textit{Chandra} and \textit{XMM-Newton} \citep{Trudolyubov} observations. For details on the depth, coverage and reliability of these catalogues we refer the reader to these papers. Of the GCs currently known to host LMXBs, 15 are included in our data. 

\subsection{LMXB relationships}

\begin{figure}
 \includegraphics[height=84mm,angle=270]{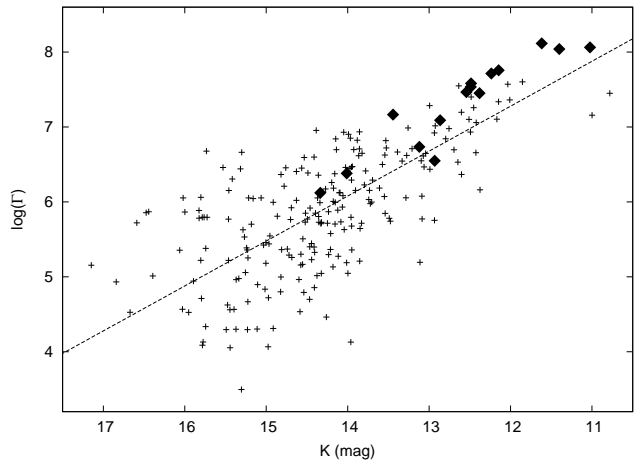}
 \caption{Stellar collision rate vs magnitude for all GCs (crosses) and LMXB hosting clusters (solid diamonds). This shows a clear relationship between the two parameters which is in good agreement with the predicted relationship $\Gamma\propto M_{tot}^{1.5}$ (dashed line). The LMXB hosting clusters are found to have higher than average collision rates for their magnitude.}
 \label{fig:lmxb}
\end{figure}

Figure \ref{fig:lmxb} shows the stellar collision rate as a function of magnitude for all GCs studied (crosses) and those containing a LMXB (diamonds). It can be seen that the LMXBs favour brighter clusters and those with higher stellar collision rates. A Kolmogorov-Smirnov (K-S) test between all GCs and the LMXB hosting GCs shows that both of these relationships are significant with probabilities of 10$^{-5}$ for luminosity and 10$^{-7}$ for collision rate that they come from the same populations. This demonstrates that the collision rates and masses of GCs are very good discriminators in selecting LMXBs. 

It can be seen from figure \ref{fig:lmxb} that there is a clear relationship between the mass of a GC and its stellar collision rate. In order to explore the relative effects of these parameters on LMXBs, we must first consider this relationship. Assuming the relationship to be linear between the logarithm of these values, we find that $\Gamma\propto M_{tot}^{1.4}$. This is consistent with the Milky Way GCs and the theoretical approximation that $\Gamma\propto M_{tot}^{1.5}$ \citep{Davies}. 

It can be seen that the LMXB hosting GCs have higher than average collision rates for their mass. To investigate whether this is a statistically significant effect, we first detrend the data using the relationship between $\Gamma$ and mass. We then run a K-S test between all GCs and the LMXB GCs and find a probability of 10$^{-4}$ that they are from the same distribution for a detrending on $\Gamma\propto M_{tot}^{1.5}$. We note that there is some uncertainty in the actual relationship between $\Gamma$ and mass due to the large scatter in figure \ref{fig:lmxb}. To ensure our results are robust to errors in this relationship, we rerun the tests assuming $\Gamma\propto M_{tot}^{1.25,1.75}$ and find probabilities of $10^{-5}$ and $10^{-3}$ respectively that they are from the same distribution. This demonstrates that, even for the steepest reasonable relationship, the LMXBs favour GCs with higher than average stellar collision rates for their mass. 

\begin{figure}
 \includegraphics[height=84mm,angle=270]{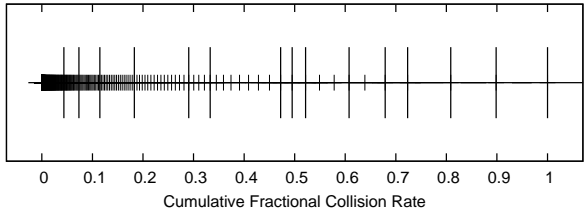}
 \caption{This shows the cumulative fractional collision rate for all GCs (small crosses) and LMXB hosting GCs (large crosses). If the formation of LMXBs is linearly proportional to GC collision rate, they should be evenly distributed along this plot.}
 \label{fig:reduced-gamma}
\end{figure}

To investigate whether the formation of LMXBs is linearly proportional to the stellar collision rate of a cluster, we follow the method of \citet{Verbunt}. First we divide all GC collision rates by the total collision rate for the whole GC system. We then sort all the clusters by this fractional collision rate and find the cumulative value for each cluster, such that the cluster values run from 0 to 1. The result of this is that the total collisions occurring in the GC system should now be evenly distributed between 0 and 1 (i.e. we expect 10$\%$ of the total collisions to occur in clusters with values in the range 0-0.1). Therefore, if the formation of LMXBs is linearly proportional to collision rate, their host clusters should be evenly spaced in this plot. It can be seen from figure \ref{fig:reduced-gamma} that the LMXB hosting clusters (large crosses) are consistent with this. 

\section{Conclusions}

We show clear relationships between the presence of a LMXB and both the stellar collision rate and mass of its host GC. 

The stellar collision rate is found to be the best discriminator in selecting LMXB hosting GCs. We suggest that the weaker relationship between mass and LMXB presence may be primarily due to the relationship between mass and stellar collision rate. Our results demonstrate that the stellar collision rate is likely to be a fundamental parameter related to the formation of LMXBs. This result is in agreement with previous studies of GCs in the Milky Way and Cen~A. 

The linear relationship found between the presence of LMXBs and stellar collision rate is in good agreement with the systems being formed by dynamical interactions. This also suggests that the current dynamical properties of the GCs are related to their current LMXB populations. 

\section*{Acknowledgements}

We would like to thank the UKIRT service team for taking these data especially Mark Rawlings and Andy Adamson. We also thank Mike Irwin for his advice regarding the WFCAM pipeline and data.

\label{lastpage}

\end{document}